\documentclass[prl,twocolumn,showpacs,amsmath,amssymb,superscriptaddress]{revtex4-1}
\usepackage{graphicx}
\usepackage{amsmath}
\usepackage{amssymb}
\usepackage{xcolor}
\usepackage{times}
\newcommand{\BibitemShut}[1]{} %handles an error due to an outdated natbib

\expandafter\def\expandafter\normalsize\expandafter{%
    \normalsize
    \setlength\abovedisplayskip{5pt}
    \setlength\belowdisplayskip{5pt}
    \setlength\abovedisplayshortskip{5pt}
    \setlength\belowdisplayshortskip{5pt}
}

\begin{document}%
%\linespread{2}\selectfont%

\newcommand*{\UNIMAINZ}{Institut f\"ur Physik, Johannes Gutenberg-Universit\"at Mainz, Staudingerweg 7, D-55128 Mainz, Germany}
\affiliation{\UNIMAINZ}
\newcommand*{\HIM}{Helmholtz-Institut Mainz, Johann-Joachim-Becherweg 36, D-55128 Mainz, Germany}

\homepage{http://www.quantenbit.de}

\title{Rydberg excitation of a single trapped ion}

\author{T.~Feldker}\email{feldker@uni-mainz.de}\affiliation{\UNIMAINZ}
\author{P.~Bachor}\affiliation{\UNIMAINZ}\affiliation{\HIM} 
\author{M.~Stappel}\affiliation{\UNIMAINZ}\affiliation{\HIM}  
\author{D.~Kolbe}\altaffiliation{Present address: Institut f\"ur Technische Physik,
Deutsches Zentrum f\"ur Luft- und Raumfahrt,
Pfaffenwaldring 38-40, D-70569 Stuttgart, Germany}\affiliation{\UNIMAINZ}\affiliation{\HIM}
\author{R.~Gerritsma}\affiliation{\UNIMAINZ} 
\author{J.~Walz}\affiliation{\UNIMAINZ}\affiliation{\HIM} 
\author{F.~Schmidt-Kaler}\affiliation{\UNIMAINZ}

\date{\today}

\begin{abstract}
We demonstrate excitation of a single trapped cold $^{40}$Ca$^+$ ion to Rydberg levels by laser radiation in the vacuum-ultraviolet at 122\,nm wavelength. Observed resonances are identified as 3d\,$^2$D$_{3/2}$ to 51\,F, 52\,F and 3d\,$^2$D$_{5/2}$ to 64\,F. We model the lineshape and our results imply a large state-dependent coupling to the trapping potential. Rydberg ions are of great interest for future applications in quantum computing and simulation, in which large dipolar interactions are combined with the superb experimental control offered by Paul traps.
\end{abstract}
 
%37.10.Ty: Trapped ions
%32.80.Ee: Rydberg states
%42.50.Ex: Optical implementations of quantum information processing and transfer

\maketitle

% Rydion Intro

The properties of Rydberg atoms are dominated by one electron being in a state of high
principal quantum number, which causes long lifetimes and large dipole
moments~\cite{reviewHaroche, reviewSaffman}.  This results in giant dipolar interactions between Rydberg
atoms~\cite{Lukin2001} which enable the formation of ultralong-range
molecules~\cite{Bendkowsky2009}, quantum logic gate operations between two neutral
atoms~\cite{Wilk2010,Isenhower2010}, and the control of the state of transmitted light
through a Rydberg sample at the single photon level~\cite{Peyronel2012}. A completely
new approach to this field of research is the Rydberg excitation of trapped
ions~\cite{Mueller2008,Schmidt2011,Li2014}, which aims to combine the long-range
Rydberg-blockade mechanism, demonstrated in the case of neutral atoms confined in
optical lattices~\cite{Amthor2010,Schauss2012,Ravets2014}, with the superb level of
control over single ions achieved in Paul traps~\cite{Schmidt2005, Roos2012}. Rydberg ions in Coulomb crystals
will allow for shaping localized vibrational modes for quantum simulation and fast
parallel execution of quantum gates~\cite{Li2013,Nath2014}. Further applications are
dynamical structural phase transitions and non-equilibrium dynamics driven by Rydberg
excitations~\cite{Li2012}.

% discussion of challenges

Two major challenges have to be met in order to access the unique features of Rydberg ions: Firstly, excitation energies are large compared to the case of neutral
atoms such that either a vacuum ultraviolet (VUV) laser
source~\cite{Schmidt2011,Kolbe2012} or multi-step excitation~\cite{HennrichPrivate} with UV lasers is required.  Secondly, the large polarizability of Rydberg states makes them very susceptible to residual electric fields in the Paul trap, where an oscillating field
with quadrupolar geometry and gradients of about 10$^7$-10$^9$~V/m$^2$ provides
stable trapping conditions. Ions are confined near the node
(field-zero) of the electric quadrupole, nevertheless residual fields at the position of the ion perturb the Rydberg state and lead to a shifted and broadened resonance.  

 %description of ion traps 

\begin{figure}
\centering
\includegraphics[width=0.45\textwidth]{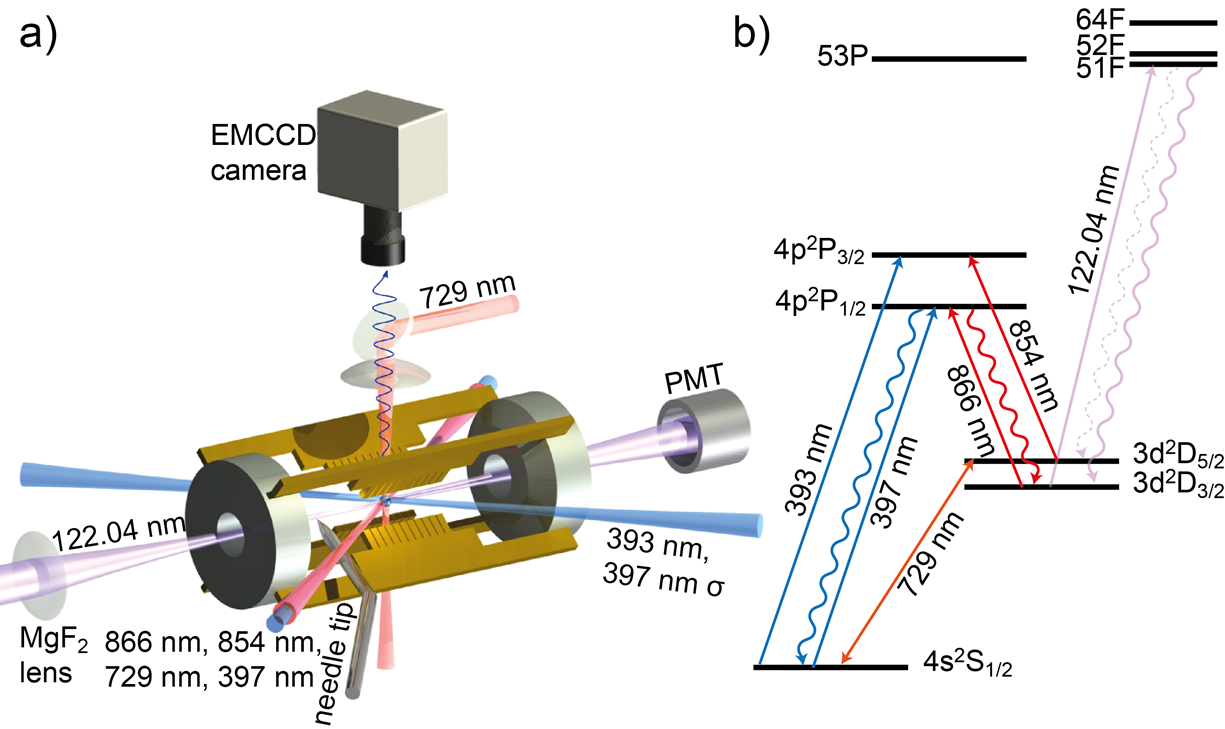}
\caption{a) Sketch of the x-shaped ion trap (B) with two segmented gold-covered DC blades and
two RF blades.  Laser beams near 397\,nm, 866\,nm, 854\,nm, 393\,nm and 729\,nm can be
switched and serve for cooling, optical pumping and state analysis.  Fluorescence
light near 397\,nm is collected by a lens with numerical aperture NA = 0.27 and imaged
on an electron-multiplying charge-coupled device (EMCCD) camera.  Holes in the endcaps provide access for the VUV beam. The VUV beam intensity is monitored with a
photomultiplier tube (PMT).  A sharp tip can be moved into the trap for aligning the
VUV beam. $~$ b) Levels and transitions in $^{40}$Ca$^+$. Ions are cooled on the
S$_{1/2}$ to P$_{1/2}$ transition near 397~nm with optical pumping on the D$_{3/2}$ to
P$_{1/2}$ transition at 866\,nm and initialized for the Rydberg excitation by optical
pumping into the D$_{3/2}$ (D$_{5/2}$) level with light near 397\,nm (393\,nm).}
\label{fig:trap}
\end{figure}

In this Letter we present laser excitation of a single trapped cold ion to Rydberg states. Ions are initialized in the metastable 3d\,$^2$D$_{3/2}$ or 3d\,$^2$D$_{5/2}$ state before they are excited to the 51\,F and 52\,F (from D$_{3/2}$) respectivly 64\,F (from D$_{5/2}$) state using vacuum ultraviolet radiation near 122\,nm. By applying a state dependent fluorescence measurement following the decay of the Rydberg state, population transfer out of the initial D-state is detected. The polarizability of the Rydberg ion is deduced from the observed line-shift and -broadening, caused by residual electric fields in the trap.

Experiments were carried out in two different linear Paul traps.  Trap (A) consists of four cylindrical rods with diameter d = 2.5\,mm at a diagonal distance of 2.2\,mm, and two
endcaps at a distance of 10\,mm~\cite{Schmidt2011,Feldker2014}.  It has a rather large trapping volume and
features stable ion trapping at very low field gradients.  Trap (B) is constructed
from x-shaped gold-covered blades at a diagonal distance of 0.96\,mm, see ref.~\cite{Jacob2014}, and Fig.~1\,a.  Axial confinement is provided by applying voltages to the
segments of the DC blades. The improved design of this trap allows for a more precise alignment of the electrodes. Thus
residual fields at the ion are reduced and the trap can be operated at higher
vibrational frequencies compared to trap (A).
%hier hin den teil über probleme mit den Fallen?
 Ions are loaded from a thermal beam of neutral calcium atoms
by photoionization with laser light near 423\,nm and 375\,nm.  Trapped ions are
illuminated by laser radiation near 397\,nm and 866\,nm for optical cooling and
fluorescence detection with an electron-multiplying charge-coupled device (EMCCD)
camera.  Additional laser beams at 393\,nm, 854\,nm and 729\,nm allow for optical
pumping, sideband spectroscopy and coherent manipulation of the 4s\,$^2$S$_{1/2}$ -
3d$^2$\,D$_{5/2}$ transition, respectively.

%desccription of VUV source

Figure 1\,b shows relevant energy levels of the calcium ion.  The Rydberg transition
is driven by VUV single-photon excitation at 122\,nm wavelength from the metastable
D$_{3/2}$ and D$_{5/2}$ levels.  To generate continuous-wave coherent VUV radiation we
employ four-wave sum-frequency mixing in mercury vapor.  Power levels in the $\mu$W
range are achieved using a triple-resonant scheme with fundmental light fields at
254\,nm, 408\,nm and 555\,nm which are generated by frequency-doubling and -quadrupling of lasers in the near infrared~\cite{Schmidt2011,Kolbe2012}. The wavelength of these lasers are monitored by a wavelength meter (HighFinesse WSU-10), which is calibrated to the 4s\,$^2$S$_{1/2}$ -
3d$^2$\,D$_{5/2}$ transition and thus accurate to about 100~MHz at the sum frequency. The VUV radiation is
focused on the ion by a pair of MgF$_2$ lenses and the transmitted power is monitored
by a photomultiplier tube (PMT). For a precise
determination of the focus we use a sharp tip which can be
temporarily moved into the trap and monitor the transmission of the VUV beam. The beam is aligned by shifting one of the MgF$_2$ lenses as well as the ion traps vacuum vessel which is mounted on a translation stage.

%Beschreibung der Suche

We search Rydberg resonances at wavelengths near 122.04\,nm, where the efficiency of
the four-wave mixing process is strongly enhanced by the nearby 7\,$^1$S-11\,$^1$P
resonance in mercury.  Inital information about the expected resonances \cite{Xu1998}
could not be reproduced, and uncertainties of theory predictions~\cite{Schmidt2011}
are too large, so that we had to search the full range of about 180\,GHz between
Rydberg states of consecutive principle quantum number.

%describtion of Rydberg excitation

The ion is excited to the Rydberg state in a sequence of steps: $~$ i)~Initialization
by optical pumping to the 3d$^2$\,D$_{3/2}$ state (lifetime: 1176(11)\,ms
\cite{Kreuter2005}). $~$ ii)~VUV excitation by a 30~ms pulse. The Rydberg state then
decays and within 30~ms is either back in the initial state, or the 3d$^2$\,D$_{5/2}$ (lifetime:
1168(9)\,ms \cite{Kreuter2005}), or the ground state 4s$^2$\,S$_{1/2}$ and the
detection efficiency is therefore dominated by the branching fractions out of the
initial state.
$~$ iii)~For the detection, ground state population is optically pumped into the D$_{5/2}$
state with resonant light at 393\,nm (90$\%$ efficiency from the branching ratio to the
D$_{3/2}$ level).  $~$ iv)~Under resonant excitation with laser light at 397\,nm and
866\,nm the ion emits no fluorecence light, if the Rydberg excitation in (ii) was
successful. Figure~2\,a shows the observed resonance.  For the data shown in
Fig.~2\,b, the pumping step (iii) is omitted in order to prove that this Rydberg state is
decaying into the D-states.

We observe Rydberg excitation to the 51\,F state at 122.041\,913(5)\,nm and to the
52\,F state at 122.032\,384(10)\,nm wavelength.  The linewidth of the excitation to
the 51\,F varies between 60\,MHz and 400\,MHz full-width at half-maximum (FWHM),
depending on the trap control parameters.  Similar measurements have been carried out
with linear three and five-ion crystals.  For the F resonances the branching ratio
between the decay from the Rydberg state into the D$_{5/2}$ and D$_{3/2}$ state is
estimated to be about 7~$\%$, using data from lower-lying Rydberg states (cf.\ NIST
atomic spectra database).  The decay of a Rydberg F-state into the S$_{1/2}$ state is
forbidden.

In addition, Rydberg excitation to the 64\,F state starting from the 3d$^2$\,D$_{5/2}$
state is observed at 122.040\,50(5)\,nm.  The resonance width is larger, about
1\,GHz, as the state is at the edge of being stable in the trap
\cite{Mueller2008,Schmidt2011}.  Consequently, the 64\,F state is not measured by
applying the electron shelving scheme as described above, but by observation of a
significantly increased ion loss rate on resonance.  Ion loss is also observed,
albeit less frequently, for the 51\,F and 52\,F resonances in $\sim 0.3\%$ of the
Rydberg excitations.

%Bild 2: a) 51F und b) 52F Resonanz, Parameter
\begin{figure}
\centering
\includegraphics[width=0.45\textwidth]{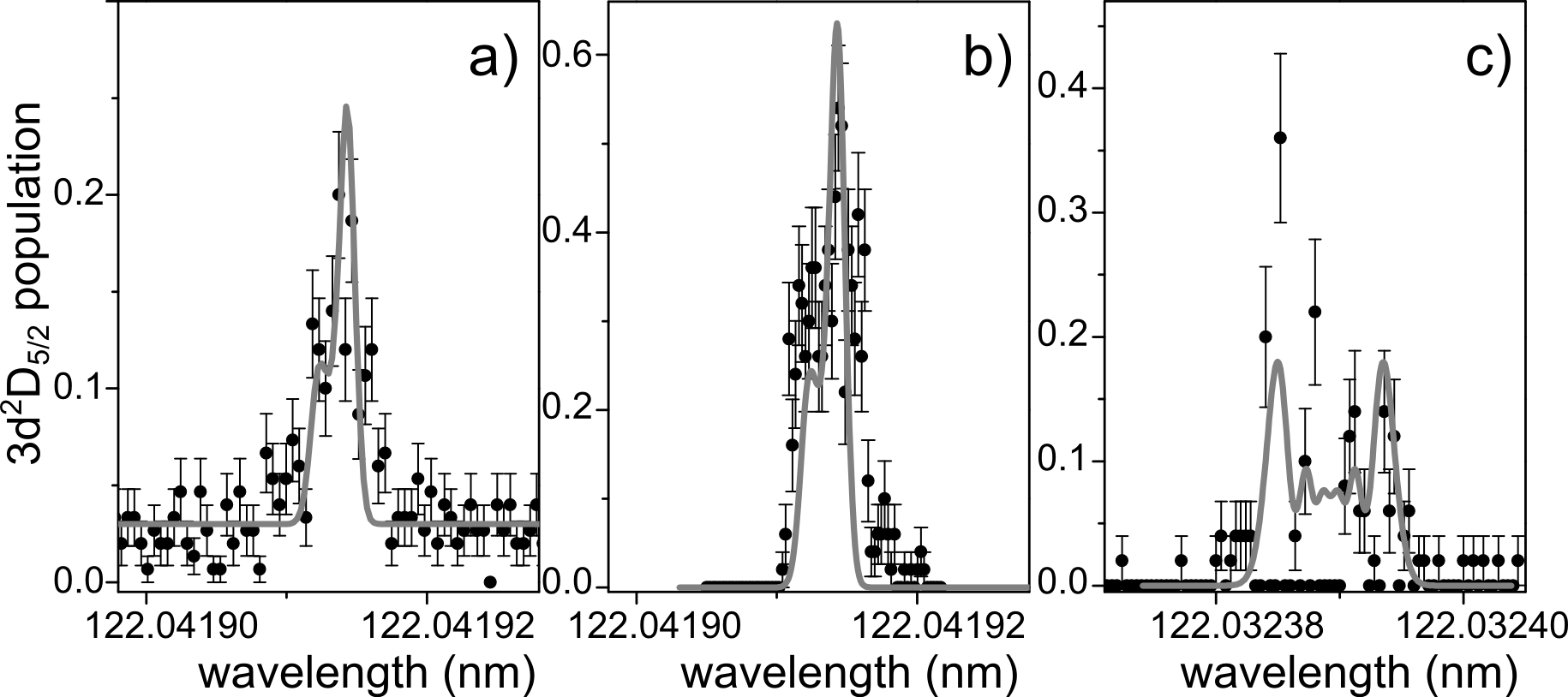}
\caption{a) and b): Rydberg excitation from the 3d$^2$\,D$_{3/2}$ to the 51\,F state, at a VUV
power of 0.5\,$\mu$W and 30\,ms pulse duration.  The trap (A) is operated at motional
frequencies of $\omega_{ax,rad}/(2\pi)= 90, 200$\,kHz. $~$ a) Ground state population
is optically pumped to the 3d$^2$\,D$_{5/2}$ state after the Rydberg excitation.  The
linewidth is $\simeq$ 60\,MHz full-width-at-half-maximum (FWHM). $~$ b) No optical
pumping, only the direct decay to 3d$^2$\,D$_{5/2}$ is observed.  This data
confirmes that the Rydberg level is a F-state (see text for details). $~$
c)~Excitation from the 3d$^2$\,D$_{3/2}$ to the 52\,F state in trap (B), operated at
motional frequencies of $\omega_{ax,rad}/(2\pi)= 270, 700$\,kHz. The line shape model
includes broadening by thermal excitation, magnetic field,
micromotion and Stark effect. 50 single shot measurements were taken for each datapoint, the errorbars depict the quantum projection noise.
}
\label{fig:Resonance}
\end{figure}

%Tabelle1: Frequenzen und Quantendefekte

\begin{table}[b]
\begin{tabular}{lllll}
\hline
\multicolumn{1}{|l|}{\textbf{state}} & \multicolumn{1}{l|}{\textbf{ wavelength from D$_{3/2}$}} & \multicolumn{1}{l|}{\textbf{wavelength from D$_{5/2}$}} & \multicolumn{1}{l|}{\textbf{$\Delta$E~cm$^{-1}$ }}  \\ \hline
\multicolumn{1}{|l|}{51F}          & \multicolumn{1}{l|}{ 122.041 913(5)~nm}          & \multicolumn{1}{l|}{--}          & \multicolumn{1}{l|}{95589.258(3)}         \\ \hline
\multicolumn{1}{|l|}{52F}          & \multicolumn{1}{l|}{ 122.032 384(10)~nm}          & \multicolumn{1}{l|}{--}          & \multicolumn{1}{l|}{95595.656(7)}               \\ \hline
\multicolumn{1}{|l|}{64F}          & \multicolumn{1}{l|}{ --}          & \multicolumn{1}{l|}{122.040 50(5)~nm}          & \multicolumn{1}{l|}{95650.901(33)}                 \\ \hline     
\end{tabular}
\caption{Levels, transition vacuum wavelengths, and energies from the S$_{1/2}$ ground state.}
\label{my-label}
\end{table}

%Identifikation der Linien als F, exakte gruende und weitere hinweise

Energy levels for Rydberg states are given by
\begin{equation}
E_{n,\delta} = -\frac{R_{\infty}}{1+\frac{m_e}{m_{\textnormal{ca}^+}}}\times \frac{Z^2}{(n-\delta)^2}
\end{equation}
where $R_\infty$ is the Rydberg constant, $m_e$ is the electron mass,
$m_{\textnormal{ca}^+}$ is the mass of the calcium ion, $Z$ is the charge-state, and
$n$ is the principal quantum number. $\delta$ denotes the quantum defect and values
from theory~\cite{Mueller2008} are $\delta_F$ = 0.026 and $\delta_P$ = 1.44 for F- and
P-states, respectively.  Measured energy differences are consistent with a nearly
vanishing quantum defect (F-state excitation) and principal quantum numbers of $n$ = 51, $n$ = 52 and $n$ =
64.  Table 1 shows the transition frequencies and energies determined.  Measured
frequencies of 411.042\,129\,776\,393\,THz for the 4s$^2$\,S$_{1/2}~$ to
3d$^2$\,D$_{5/2}$ transition~\cite{Chwalla2009} and 1.819\,599\,021\,504\,THz for the
3d$^2$\,D$_{5/2}$ - 3d$^2$\,D$_{3/2}$ fine structure splitting~\cite{Yamazaki2008}
have been used in the data analysis. The
identification of the resonances is supported by the fact that decay from the Rydberg
F-states is mainly into the 3d\,$^2$D states as observed experimentally (see
Fig.~2\,b), in contrast to Rydberg P-states which would predominatly decay into the
4s$^2$\,S$_{1/2}$ state~\cite{LiPrivate}.

%Even more, polarizability $\alpha$ = 4$\pi\times$ 500~MHz/(V/m)$^2$ determined from
%the shape of the resonance is in good agreement with the numerical obtained value of
%$\alpha_F$ = 4$\pi\times$ 400~MHz/(V/m)$^2$ in contrast to $\alpha_P$ = 4$\pi\times$
%-40~MHz/(V/m)$^2$.  evtl. den letzten Satz hier weglassen und im n???chsten
%Abschnittt bringen?

%messungen und rechnungen zu B Feld, zu Temperatur und zu MM kompensation als basis f??r linienmodell

In order to model the observed resonance lines, the electric field, the motional state of
the ion and the magnetic field are taken into account: The electric field at the position of the ion
is determined from the micromotion line-broadening of the 4s\,$^2$S$_{1/2}$ --
4p\,$^2$P$_{3/2}$ transition.  Due to fabrication imperfections in the geometry of
trap (A), the ion is exposed to an alternating electric field proportional to the trap
drive amplitude $E_{Ion} \propto U_{RF}$ $\times$ sin($\Omega$t), where $U_{RF}$ is the
applied radiofrequency voltage at the trap-driving frequency $\Omega/2 \pi$.
This causes micromotion in the axial direction along the VUV laser beam, which leads
to a modulation of the Rydberg energy due to the electric polarizability $\alpha$, and
--in addition-- to a modulation of the Doppler effect. The instantenous resonance frequency reads:

\begin{equation}
\omega (t) = \omega_0 + k x_{mm}\Omega\sin \Omega t-\frac{\alpha E_{ion}^2}{2}\cos^2\Omega t.
\end{equation}

\noindent Here, $\omega_0$ is the bare resonance frequency, $k$ is the wavenumber of the laser along the motion of the ion, $E_{ion}$ is the electric field amplitude
and $x_{mm}$ is the micromotion amplitude. From this equation we can derive the
temporal shape of the resonance laser field

\begin{eqnarray}
E_{res} (t) &\propto&  e^{-i\omega_0t}e^{i2\beta_{\alpha}\Omega t}\hspace{3mm} \nonumber \\ 
&\times& \sum_n J_n\left(\beta_{mm}\right)e^{in(\Omega t +\frac{\pi}{2})} \hspace{3mm}  \nonumber \\
&\times& \sum_m (-1)^m J_m\left(\beta_{\alpha}\right)e^{2im\Omega t }. \label{eq_Emod}
\end{eqnarray}

\noindent In this equation we define the modulation index in the Bessel functions
$J_n(\beta)$ due to the axial micromotion by $\beta_{mm}=k x_{mm}$ and due to the
oscillating Stark shift by $\beta_{\alpha}=\alpha E_{ion}^2/8\Omega$. The second
exponent in Eq.~(\ref{eq_Emod}) denotes a static frequency shift due to the fact that
the modulated Stark shift can only reduce the transition frequency to the F states. It
is apearant that the lineshape comes from a complicated interplay from the sidebands
at $n \times$ $\Omega_{rf}$ caused by the micromotion and the sidebands at $2n\times$
$\Omega_{rf}$ caused by the oscillating Stark shift. Thermal ion motion at T $\approx$ 5\,mK and the
magnetic field of B = 0.45\,mT, as determined from sideband
spectroscopy of the 4s\,$^2$S$_{1/2}$ - 3d\,$^2$D$_{5/2}$ transition, are leading to a broadening of $\approx$ 10~MHz,
preventing us from resolving the sidebands.
  
In trap (B), residual electric fields are generated by charging the surface of the trap
electrodes with VUV radiation. The resulting field is thus pointing in the radial
direction and linebroading by micromotion parallel to the axial aligned VUV beam is
negligible while the Stark effect remains the dominant line broadening effect.

With the known electric field amplitude $E_{Ion}$ we determine $\alpha$ from the fit of the observed
resonance line shape, the data is presented in Fig.~3~a. Calculated lineshapes at various electric fields for the transitions to the 51F- and 51P-state are presented in Fig.~3~b and c, the polarizabilities used in the calculation are estimated from second-order perturbation theory whith
neglected spin-orbit coupling. We use the quantum defects found
in~\cite{Mueller2008} and the method developed in~\cite{Kostelecky1985} to obtain
analytical approximate wavefunctions for computing matrix elements. As a check to the
accuracy of this approach, we compare the calculated lifetime of the 4P state of
6.55~ns, to the experimental value of 7~ns, which agrees reasonably well. In the
second-order perturbation theory, we set $m_L=0$, which is justified as this state
experiences the largest Stark shift for the states of interest. In the calculation, we
take couplings to the states $n=40,..,60$ into account, which was found to lead to
convergence.

The experimentally determined $\alpha/2 = 5\times$10$^2$\,MHz/(V/cm)$^2$ is close to the
polarizability of the 51\,F state, $\alpha_{51F}/2 \approx
4\times$10$^2$\,MHz/(V/cm)$^2$ obtained from second-order pertubation theory and
neglecting spin-orbit coupling.  Under the assumption that the
Rydberg level is the 51\,P state the determined electric polarizability would be
inconsistent with the theoretical value of $\alpha_{51P}/2 \approx
-37\times$\,MHz/(V/cm)$^2$ from ref.~\cite{Kamenski2014}.  Again, this supports
the identifiction of the Rydberg levels as F-states.

%diskussion der polarisierbarkeit von F und P linien

%Bild 3: Modellierung der Linienform f??r die a) F und die b) P Resonanz und c) die gemessene Verschiebung der Linie gegen den RF Spannung
\begin{figure}
\centering
\includegraphics[width=0.45\textwidth]{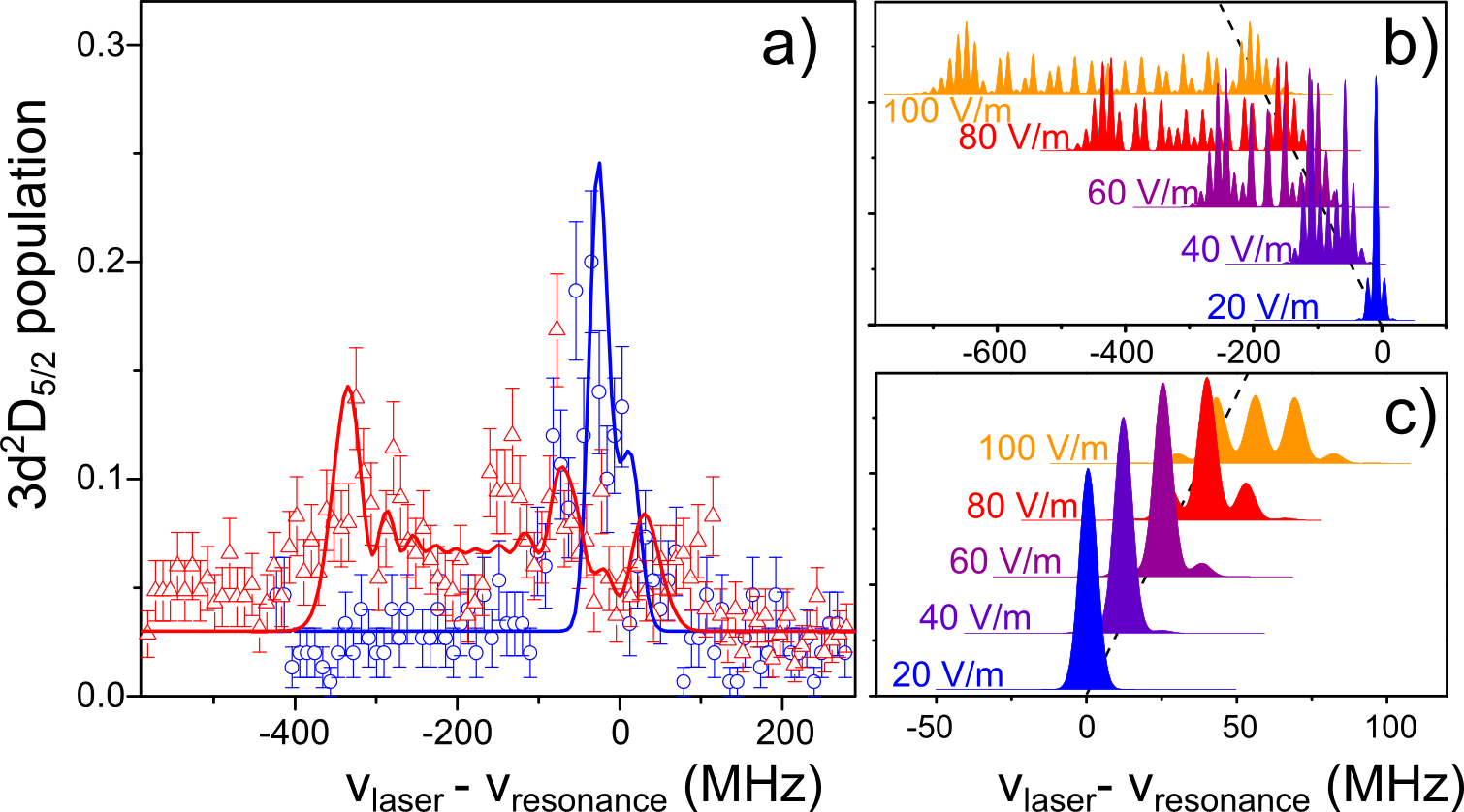}
\caption{a) Experimental data (trap A) of the 3d\,$^2$D$_{3/2}$ to 51\,F resonance with $E_{Ion}$ =
24\,V/m (blue) and 84\,V/m (red).  Fitting the data with our lineshape model, we
determine $\alpha_{51F}$/2 = 5$\times$10$^2$\,MHz/(V/cm)$^2$ for the
polarizability. 50 single shot measurements were taken for each datapoint, the errorbars depict the quantum projection noise. $~$ b) Calculated lineshapes for the D$_{3/2}$ to 51\,F transition at
different residual electric fields.  The axial micromotion is neglected (trap B) and
cooling to the Doppler limit is assumed.  Features of this calculation are compatible
with experimental data. $~$ c) Set of lineshapes for the transition D$_{3/2}$ to 51\,P
with the same parameters as in b).  Features of this calculation are incompatible with
experimental data. This observation corroborates the indentification of the excited Rydberg levels as F-states. 
}
\label{fig:LineShape}
\end{figure}

%weitere schritte im experment darstellen: linien mit niedrigem n ca 20 suchen, verweis auf Markus Hennrich exp.., P linien suchen, h??here rabifrequenz mit engere fokussierung..rydberg blockade
In both traps we have observed that Rydberg excitation becomes unstable after several months
of experimenting. We conjecture surface contamination (e.g. calcium from the loading atomic beam) and VUV generated photo electrons to be the reason for this observation and plan to use a dedicated ion-loading zone well separated from the
spectroscopy zone~\cite{Home2009,Ruster2014} to avoid
VUV-induced photoemission of electrons and charging of electrodes.
In the future increased VUV power and tighter focussing will allow for significantly higher excitation rates.  This should enable the
excitation of Rydberg P-states, where the dipole matrix elements are lower by about one order of magnitude.  Much narrower excitation
resonances are expected from Rydberg P-states, because the electric polarizability is
lower, $\alpha_P/\alpha_F \simeq$ 0.1.  For a single ion in the motional ground
state, and exposed to residual electric field of 65~V/m as measured from the Rydberg
resonance in Trap (B), we expect a modulation index of 0.6. This results in a strong
carrier and resolved sidebands at a frequency of $\pm$2$\times \Omega_{rf}$ weaker by
a factor of three.
In view of driving coherent dynamics, we note that already with the achieved values
for VUV power of 3\,$\mu$W and a focussing of 10\,$\mu$m, Rabi frequencies of up to
2$\pi \times$ 150\,kHz for the transition 3d\,$^2$D$_{3/2}$ to 52\,P are
expected~\cite{Schmidt2011}, which is much faster than the natural decay of
217\,$\mu$s~\cite{Glukhov2013}.

%\paragraph{Conclusion.}

In conclusion we have observed and identified Rydberg excitations of a single cold ion
and measured the lineshape of the resonance which implies a strong state-dependent coupling to the electric trapping potential. 
This experimental work is a starting point for establishing a
novel platform of Rydberg matter, where the unique possibilities of state dependent electric forces 
as well as long-range dipole-dipole interactions are combined with the outstanding
control over quantum states in trapped ion cyrstals.

\paragraph{Acknowledgments} The authors acknowledge helpful discussions with W. Li, I. Lesanovsky, P. Zoller and M. Drewsen. We acknowledge G. Jakob and S. Wolf for support with the assembly of trap (B). This work was funded by the BmBF and the chist-era network(R-ION consortium) and by the European Union H2020 FET Proactive project RySQ (grant N. 640378).

%%%%%%%%%%%%%%%%%%%%%%%%% Bibliography

\bibliographystyle{apsrev4-1}

\end{document}